# On-Chip Optical Switching with Epsilon-Near-Zero Metamaterials

**Yu Peng** [1,2*]

[1] *School of Engineering and Applied Sciences, Harvard University, 9 Oxford Street, Cambridge, Massachusetts 02138, USA.*
[2] *Research Center for Advanced Science and Technology, The University of Tokyo, Tokyo, Japan*

[*] *peng@g.ecc.u-tokyo.ac.jp*
[*] *harvardpeng@gmail.com*

Abstract: We present a multifunctional on-chip optical device utilizing epsilon-near-zero (ENZ) metamaterials, allowing precise beam control through phase modulation. This design acts as both an all-optical switch and a tunable beam splitter, providing compact and scalable solutions for integrated photonic applications.

## 1. Introduction

In photonic circuits, the ability to precisely control wave propagation is crucial for devices like switches, filters, and beam splitters. ENZ metamaterials offer significant advantages by enabling phase-free propagation, making them ideal for applications such as supercoupling, beam steering, and enhanced nonlinear effects.

In this paper, we present an on-chip, in-plane optical switch operating in the telecom wavelength range, designed using ENZ metamaterials. The device consists of a square array of low-aspect-ratio silicon pillars on a silicon-on-insulator (SOI) substrate, embedded in an SU-8 slab waveguide. We demonstrate that one beam within the ENZ structure can be fully controlled by the phase of a second modulator beam. These two coherent beams are launched simultaneously from both sides of the ENZ metamaterial input (Figure 1).

This method can be easily extended to other frequency ranges by adjusting the dimensions of the metamaterials or using alternative ENZ materials. Our analytic calculations align closely with numerical simulations, validating the design's robustness. The principles and techniques presented here are applicable to a wide range of nanophotonic devices and components, including filters, sensors, and thermal emitters.

## 2. Theory and Design

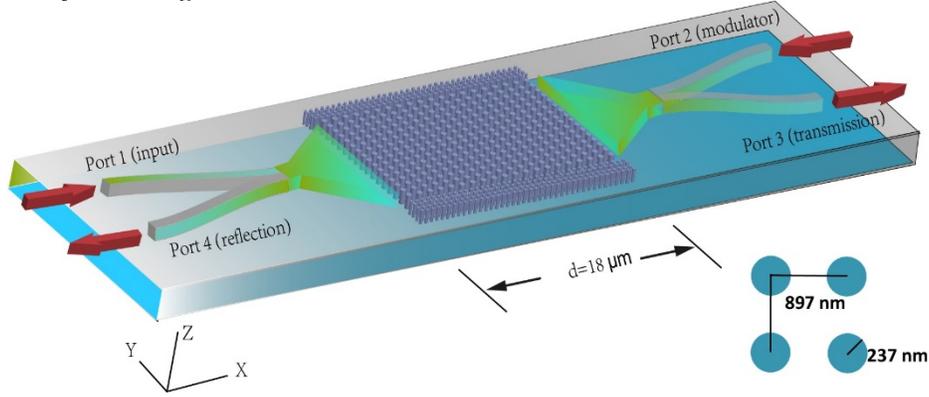

Figure 1. Schematic of the on-chip optical switch showing the interaction of input beams from ports 1 and 2 with the ENZ metamaterial, leading to tunable transmission at port 3 and reflection at port 4.

The waveguide consists of an ENZ metamaterial with a pitch $a = 897$ nm, a radius of $r = 237$ nm and a height of $h = 831$ nm, just shown in the inset at the right-bottom corner of figure 1. The total length of the device is $d = 18$ μm. Different optical properties such as the Dirac-cone like photonic bandstructure, the effective refractive index $n_{eff}$, the effective permittivity $\varepsilon$ and permeability $\mu$ or the propagation losses are described in supplemental materials (supplementary part a). The optical system consists of two input ports; one on the left of the medium (Port 1) and the other on the right (Port 2). Port 1 and port 4 have the input and output fields $E_1$ and $O_1$ in the $+x$ and $-x$ directions, respectively, in the non-absorbing incident medium. Similarly, port 2 and port 3 have the input and output fields $E_2$ and $O_2$ in the $-x$ and $+x$ directions, respectively (Figure 1). Those definitions allow us to express the relation between input and output fields using a scattering matrix formalism as in 23-27. By taking into account the amplitude ratio and phase difference between the two input beams, outputs can be written as

$$\begin{bmatrix} O_1 \\ O_2 \end{bmatrix} = \begin{bmatrix} \rho_1 & t_2 \\ t_1 & \rho_2 \end{bmatrix}\begin{bmatrix} E_1 \\ E_2 \end{bmatrix} \begin{bmatrix} O_1 \\ O_2 \end{bmatrix} = \begin{bmatrix} r_1 & t_2 \\ t_1 & r_2 \end{bmatrix}\begin{bmatrix} E_1 \\ E_2 \end{bmatrix} \qquad (1)$$

For each medium, the TE and TM waves are presented in 26-27, where the reflection coefficient from port 1 and port 2 are defined as $\rho_1 = |\rho_1|e^{i\varphi_{\rho 1}} \rho_1 = |\rho_1|e^{i\varphi_{\rho 1}}$ and $\rho_2 = |\rho_2|e^{i\varphi_{\rho 2}} \rho_2 = |\rho_2|e^{i\varphi_{\rho 2}}$. The reflection phase-change from port 1 and port 2 are defined as $\varphi_{\rho 1}$ and $\varphi_{\rho 2}$. $K_1 K_1$ and $K_2 K_2$ are the transverse wavevector in the metamaterials. The transmission coefficient through port 1 and port 2 are defined as $\tau_1 = |\tau_1|e^{i\varphi_{\tau 1}} \tau_1 = |\tau_1|e^{i\varphi_{\tau 1}}$ and $\tau_2 = |\tau_2|e^{i\varphi_{\tau 2}} \tau_2 = |\tau_2|e^{i\varphi_{\tau 2}}$, where the transverse phase-change through port 1 and port 2 are defined as $\varphi_{\tau 1} = K_1 d \varphi_{\tau 1} = K_1 d$ and $\varphi_{\tau 2} = K_2 d \varphi_{\tau 2} = K_2 d$. Since port 1 and port 2 are symmetric, the transmission and reflection through the coefficients of port 2 and port 1 are the same, so $|\tau_2| = |\tau_1| = \tau \tau_2 = \tau_1 = \tau$, $|\rho_1| = |\rho_2| = |\rho| \rho_1 = \rho_2 = \rho$ and $\varphi_{\tau_1} = \varphi_{\tau_2} = Kd$. Consequently, we have

$$\begin{bmatrix} O_1 \\ O_2 \end{bmatrix} = \begin{bmatrix} |\rho|e^{i\varphi_\rho} & |\tau|e^{i\varphi_\tau} \\ |\tau|e^{i\varphi_\tau} & |\rho|e^{i\varphi_\rho} \end{bmatrix}\begin{bmatrix} |A|e^{i\varphi_1} \\ |A|e^{i\varphi_2} \end{bmatrix}\begin{bmatrix} O_1 \\ O_2 \end{bmatrix} = \begin{bmatrix} |\rho|e^{i\varphi_\rho} & |\tau|e^{i\varphi_\tau} \\ |\tau|e^{i\varphi_\tau} & |\rho|e^{i\varphi_\rho} \end{bmatrix}\begin{bmatrix} |A|e^{i\varphi_1} \\ |A|e^{i\varphi_2} \end{bmatrix} \qquad (2)$$

The initial electric fields of port 1 and port 2 are $E_1 = |A_1|e^{i\varphi_1} E_1 = |A_1|e^{i\varphi_1}$ and $E_2 = |A_2|e^{i\varphi_2} E_2 = |A_2|e^{i\varphi_2}$, where $\varphi_1$ and $\varphi_2$ are the initial value of phase of $E_1$ and $E_2$ respectively, and we assume that $A_1 = A_2 = A$ is the amplitude of $E$. Therefore, the output of port 1 is

$$O_1 = |\rho_1|e^{i\varphi_{\rho_1}}|A|e^{i\varphi_1} + |\tau_2|e^{i\varphi_{\tau_2}}|A|e^{i\varphi_2} \tag{3}$$

and the output of port 2 is

$$O_2 = |\tau_1|e^{i\varphi_{\tau_1}}|A|e^{i\varphi_1} + |\rho_2|e^{i\varphi_{\rho_2}}|A|e^{i\varphi_2} \tag{4}$$

Adding initial value of phase of E field from port 1 $\varphi_1=0$ and phase change of reflection $\varphi_\rho=0$, it yields

$$\begin{bmatrix}O_1\\O_2\end{bmatrix} = \begin{bmatrix}|\rho| & |\tau|e^{iK_fd}\\|\tau|e^{iK_fd} & |\rho|\end{bmatrix}\begin{bmatrix}|A|\\|A|e^{i\varphi_2}\end{bmatrix} \tag{5}$$

The output of port 4 is the interference of reflection of port 1 and transmission of port 2. The output of port 3 is the interference of reflection of port 2 and transmission of port 1. From Eq. (3) and Eq. (4), this phenomenon can be expressed with the formulas below

$$I_{total} = I_1 + I_2 + 2\sqrt{I_1 I_2}\cos\delta \tag{6}$$

$$\delta = (K_1 - K_2)r + (\varphi_1 - \varphi_2) \tag{7}$$

Where $I_{total}$ is interference, $I_1 = |\rho_1||A|$ is intensity of the input reflection beam, $I_2 = |\tau_2||A|$ is intensity of the counter-propagating modulation beam and $\delta$ is the phase difference between the two beams.

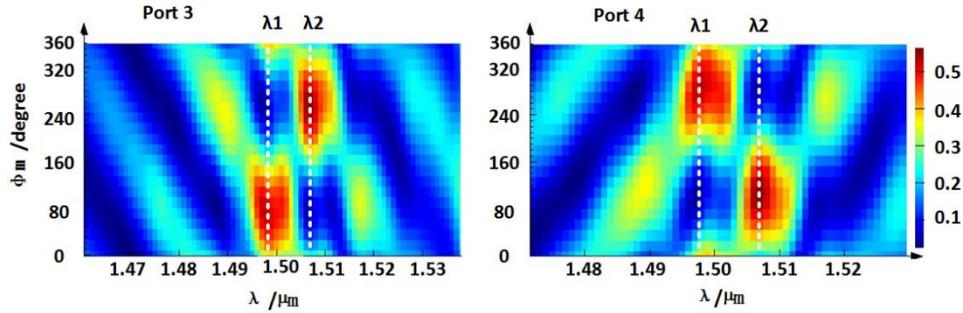

Figure 2. Transmission (output of port 3) and Reflection (output of port 4) spectra with varying phase of the modulator $\phi_m$. $\lambda_1=1498$ nm, $\lambda_2=1507$ nm.

If we apply $Kd = \frac{2\pi n_{eff}}{\lambda}d = m\pi$ to our ENZ platform at $\lambda_1=1498$ nm (TM wave) and the effective index of a propagating ENZ wave is $n_{eff} = 0.187603$. We find that in the transverse direction the thickness of 18 μm obtained from the admittance matching is in good agreement with the $m = 4.5$ (for specific calculation, see supplementary part b), implying that ENZ occurs due to critical coupling to a propagating wave in the longitudinal direction along the metamaterials. Here, we use this ENZ for on-chip optical switching since ENZ can be designed for any wavelength as needed. Moreover according to $Kd = \frac{2\pi n_{eff}}{\lambda}d = m\pi$, the length of chip can be designed very small since the effective index of a propagating ENZ wave is very small, $n_{eff} = 0.187603$.

Similarly, if we apply equations to this ENZ at $\lambda_2= 1507$ nm (TM wave), and the effective index of a propagating ENZ wave is $n_{eff} = 0.147188$. We find that in the transverse direction the thickness of 18 μm obtained from the admittance matching is in good agreement with the $m = 3.5$. ENZ occurs due to critical coupling to a propagating wave in the longitudinal direction. Theoretical output intensities as the relative phase (which equals the modulator phase $\phi_m$ when initial phase of input is set to zero ) of the input beams is varied, showing intensities emitted to the right and left sides of the slab. We get the simulation with Lumerical FDTD solution 19-20, shown in figure 2.

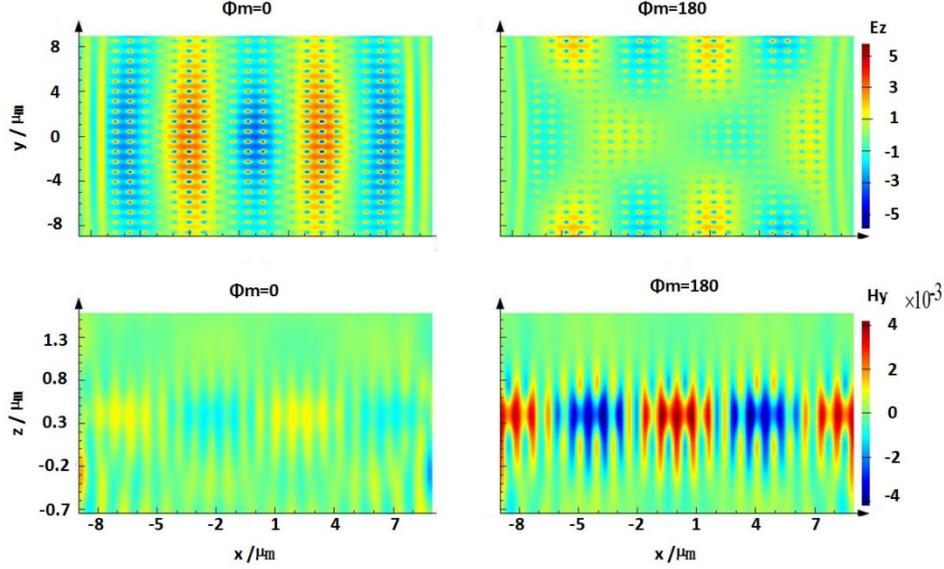

Figure 3. (Top left) Profile of in-plane component of electric field $E_z$ at $\phi_m = 0$ degree at 1498 nm; (Bottom left) Profile of out-plane component of the magnetic field $H_y$ at $\phi_m=0$ degree for 1498 nm; (Top right)Profile of in-plane component of electric field $E_z$ at $\phi_m=180$ degree of 1498 nm; (Bottom right) Profile of out-plane of $H_y$ at $\phi_m=180$ degree at 1498 nm.

We investigate now the loss distribution as a function of the relative phase shift ϕ and incident wavelength λ (supplementary part c). In principle, the loss distribution can be coherently controlled by the phase difference between the two input beams. The loss distribution versus phase 27-29 can be expressed as

$$I = I_0 \sin^2\left(\frac{\phi}{2}\right) \quad I = I_0 \sin^2\left(\frac{\phi}{2}\right) \qquad (8)$$

where $\phi\phi$ is the phase-difference of these two beams, and $I_0 I_0$ is the initial intensity when the phase-difference equals zero. To simplify the calculation, we assume the phase of the input beam is $\varphi_1=0$ and the phase of the modulator beam is $\phi_m$ can be tuned. When the difference of phase of these two beams equals 180 degree, the electric fields interferes destructively, but the magnetic fields add (or vice versa), so that energy shifts from one type of field to another in the same region (Figure 3). when the electric field is completely cancelled, the magnetic field is added up. In order to squeeze the effect in our case we design our optical device by controlling the phase of modulator around 90 degree and around 270 degree, which is far from the area of 0 degree and the area of 180 degree.

We will take full advantage of those well-known interference effects to design several applications such as an optical switch or a tunable beam splitter.

3. Optical switch

As a first application, we illustrate the behavior of our ENZ platform as a switch. In this configuration, a single beam of light (from left to right) acts as input to the metamaterial. The second beam of light (from right to left) enters the metamaterial simultaneously and acts as modulator. When the phase of the modulator is 90 degree out-of-phase compared to the input beam, the switch is in the "on" state. The output of port 3 is high (Figure 4). The switch can be turned to the "off" state, when the phase of the modulator is 270 degree. As a result, the output of port 3 can vary between "on" and "off" states (Figure 4). The threshold is set to 40%, and the logic operation is summarized in Table 1. The bandwidth is less than 10 nm.

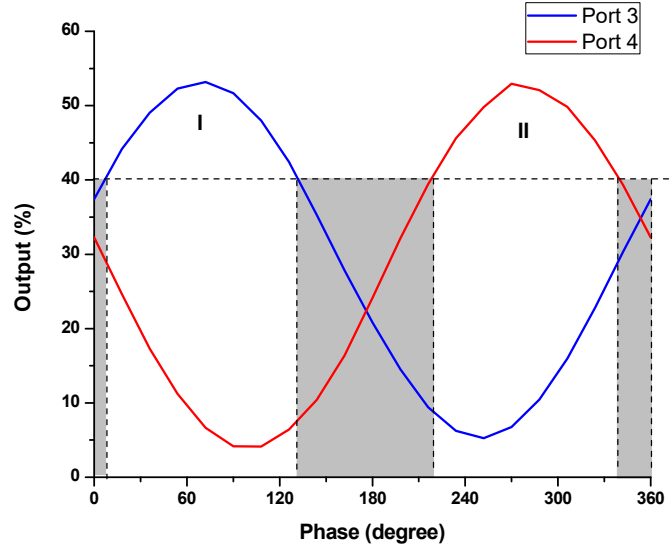

Figure 4. Output channels versus phase of modulation. The gray area corresponds to outputs with an intensity lower that the 40% threshold.

| Operation | Wavelength [nm] | Phase of port1(Input) [°] | Phase of port 2(Modulator) [°] | Port 3 (out) | Port 4 (out) |
|---|---|---|---|---|---|
| Switch | 1498 | 0 | 90 | ON | OFF |
| Switch | 1498 | 0 | 270 | OFF | ON |

## 4. Tunable beam splitter

A second application is our ENZ platform acting as a tunable beam splitter. Once again, the input (left to right) and modulator (right to left) beams are sent simultaneously. Here, the input beam should have a bandwidth of 20 nm, centered around the ENZ wavelength. When the phase of the modulator beam is 90 degree, at port 3 we get the output of $\lambda_1$ (Figure 2). At port 4, we get the output of $\lambda_2$. By changing the phase of the modulator to 270 degree, we obtain the opposite situations at port 3 and port 4. As a result, the beam splitter can be tuned by changing the phase of modulator beam (Table 2). The tunable beam splitter phenomena are summarized in Table 2.

| Beam splitter | Port1(input) | Port2(modulation) | Port3(out) | Port4(out) |
|---|---|---|---|---|
| Status 1 | 0 degree | 90 degree | On (1498nm) | On (1507nm) |
| Status 2 | 0 degree | 270 degree | On (1507nm) | On (1498nm) |

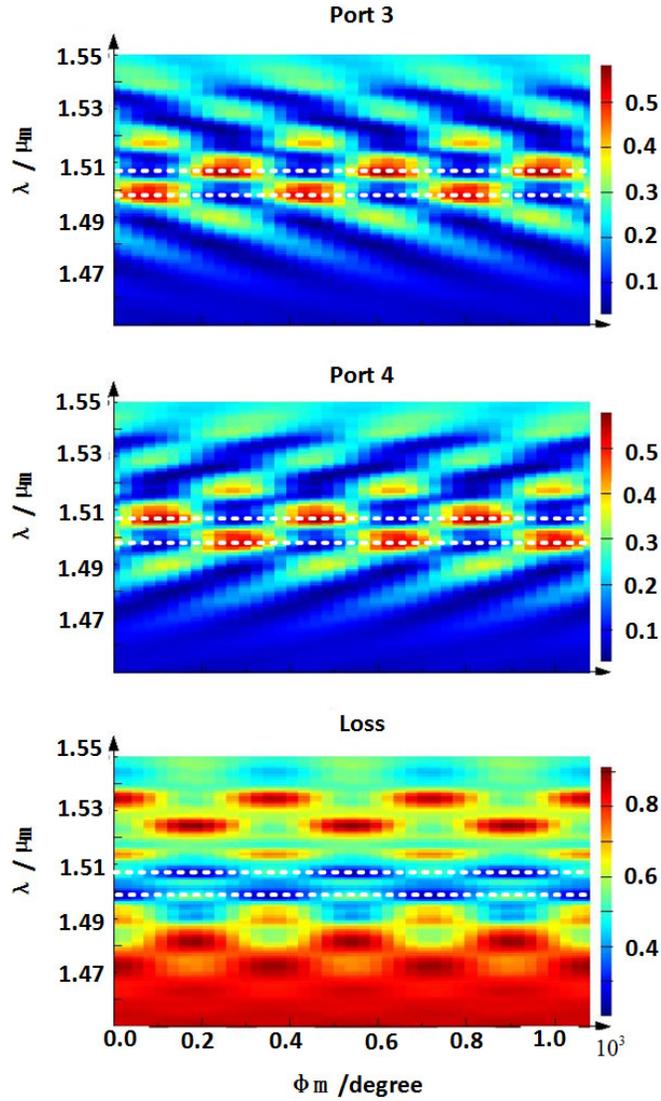

Figure 5. Transmission (top), reflection (middle) and loss distribution (bottom) versus phase of the modulator beam. White lines correspond to the two operation wavelengths.

We investigate the port 3 and port 4 as a function of the relative phase shift $\phi_m$ and incident $\lambda$ (Figure 5). It shows we have many options to make those ports in the ON and OFF states by changing the phase of modulator. Finally, we investigate the loss distribution as a function of the relative phase shift $\phi_m$ and incident wavelength. It shows there are obvious loss change around 0 degree, 180 degree, 360 degree of phase of modulator etc., which actually have impact to our device. So we have to choose the phase of modulator far from those points to squeeze the impact, such as 90 degree, 270 degree, and 450 degree etc.

## 5. Conclusion

We have successfully demonstrated a multifunctional optical platform based on ENZ metamaterials, capable of optical switching and beam splitting. Future research will focus on further miniaturization and integration with silicon photonics, with the goal of enhancing performance across broader wavelength ranges and expanding its applicability to advanced photonic systems.

## Author Contributions

Yu Peng and Michaël Lobet conceived the basic idea for this work. Yu Peng carried out the FDTD simulations. Michaël Lobet, Sarah Griesse Nascimento, Haoning Tang analysed the results. Eric Mazur and Michaël Lobet supervised the research and the development of the manuscript. Yu Peng wrote the first draft of the manuscript, and all authors subsequently took part in the revision process and approved the final copy of the manuscript.

## Acknowledgments

The authors thank Anna Shneidman, Linbo Shao, and Boris Desiatov of Harvard University for discussions, Yang Li of Tsinghua University for assistance with finite-difference time-domain simulations. M. L. is a recipient of a fellowship of the Belgian American Educational Foundation.